\documentclass[%
 reprint,
 amsmath,amssymb,
 aps,showkeys,https://www.overleaf.com/project/608bd335d666745eb22d6f68
]{revtex4-2}

\usepackage{xcolor}

\usepackage[ruled]{algorithm2e}
\usepackage{siunitx}
\usepackage{graphicx}
\usepackage{dcolumn}
\usepackage{bm}
\usepackage{hyperref}
\usepackage{multirow}
\usepackage{diagbox}
\usepackage{tikz}
\usetikzlibrary{shapes, arrows}
\usepackage[bottom]{footmisc}
\usepackage{xcolor}
\hypersetup{
    colorlinks,
    linkcolor={blue!60!black},
    citecolor={blue!60!black},
    urlcolor={blue!60!black}
}

\DeclareMathOperator*{\argmax}{\arg\!\max}

\tikzstyle{startstop} = [rectangle, rounded corners, 
minimum width=3cm, 
minimum height=1cm,
text centered, 
draw=black, 
fill=red!30]

\tikzstyle{io} = [trapezium, 
trapezium stretches=true, 
trapezium left angle=70, 
trapezium right angle=110, 
minimum width=3cm, 
minimum height=1cm, text centered, 
draw=black, fill=blue!30]

\tikzstyle{process} = [rectangle, 
minimum width=3cm, 
minimum height=1cm, 
text centered, 
text width=3cm, 
draw=black, 
fill=orange!30]

\tikzstyle{decision} = [diamond, 
minimum width=3cm, 
minimum height=1cm, 
text centered, 
draw=black, 
fill=green!30]
\tikzstyle{arrow} = [thick,->,>=stealth]
\tikzset{
    rectangle connector/.style={
        connector,
        to path={(\tikztostart) -- ++(#1,0pt) \tikztonodes |- (\tikztotarget) },
        pos=1.5
    },
    rectangle connector/.default=-2cm,
    straight connector/.style={
        connector,
        to path=--(\tikztotarget) \tikztonodes
    }}

\begin{document}

\title{Global Analysis of LISA Data with Galactic Binaries and Massive Black Hole Binaries}

\author{Stefan H. Strub}
 \email{stefan.strub@erdw.ethz.ch}
\author{Luigi Ferraioli}%
\author{Cédric Schmelzbach}%
\author{Simon C. Stähler}%
\author{Domenico Giardini}%
\affiliation{%
Institute of Geophysics, ETH Zurich\\ Sonneggstrasse 5, 8092 Zurich, Switzerland
}%

\begin{abstract}
The Laser Interferometer Space Antenna (LISA) is a planned space-based observatory to measure gravitational waves in the millihertz frequency band. This frequency band is expected to be dominated by signals from millions of Galactic binaries and tens of merging massive black hole binaries. The LISA Data Challenge 2a is focused on robust signal extraction from a blend of these two types of gravitational wave signals. Here, we introduce a novel high performance and cost-effective global fit pipeline extracting and characterizing galactic binary and massive black hole binary signals and estimate the noise of the residual. We perform the pipeline in a time-evolving weekly analysis starting with an observation time of $\SI{1}{week}$ until we reach a full year. As expected we detect more galactic binaries and massive black hole binaries bringing the noise estimate of the residual closer to the instrument noise with each week of additional observation time. Furthermore, we present a novel maximum likelihood estimate-based algorithm for extracting multiple massive black hole binaries. Additionally, we demonstrate a massive black hole binary signal extraction with a more accurate LISA response, considering higher harmonic modes, in a noisy data set.


\end{abstract}

\keywords{Gravitational Waves, LISA, Galactic Binaries, Massive Black Hole Binaries, LISA Data Challenge, Global Fit}
\maketitle

\section{Introduction}
The detection of gravitational waves (GWs) by the LIGO detector in 2015 marked a significant breakthrough in astrophysics \cite{abbott2019gwtc}. This breakthrough and the successful technology demonstration with LISA Pathfinder paved the way for LISA, a space-borne observatory designed to observe low-frequency GWs in the range from $\SI{0.1} {\mathrm{mHz}}$ to $\SI{1} {\mathrm{Hz}}$ \cite{lisaredbook}.

In the LISA frequency band, the most common sources are tens of millions of Galactic binaries (GBs), primarily white dwarf pairs, emitting quasi-monochromatic GWs. Unlike merging black holes, GBs remain separate throughout their lifetimes, allowing us to observe GBs with LISA continuously. It is estimated that the analysis of the LISA data will reveal tens of thousands of these interfering signals, with the remainder forming a galactic foreground noise \cite{digman2022lisa}. By precisely measuring the parameters of the GBs, valuable insights into the evolution of binary systems will be gained \cite{taam1980gravitational, willems2008probing, nelemans2010chemical, littenberg2019binary, piro2019inferring}.

In addition to signals from GBs, LISA will also detect GWs from massive black hole binaries (MBHBs) \cite{klein2016science}. Unlike the GBs' quasi-monochromatic GW signals, MBHBs mergers with total detector-frame masses ranging from $\SI{10^5}{\textup{M}_\odot}$ to $\SI{10^7}{\textup{M}_\odot}$ are recorded as a chirp with a high signal-to-noise ratio (SNR) making them especially prominent in the data collected  \cite{katz2022fully}. Estimates suggest LISA could detect 1 to 20 MBHB mergers annually \cite{berti2016spectroscopy, katz2020probing}, offering a window into fundamental physics, cosmology, and astrophysics \cite{petiteau2011constraining, gair2013testing, bogdanovic2014supermassive, klein2016science}.

Analyzing the LISA data presents a difficult task due to the overlap of numerous signals in the same data stream. To address this challenge, the LISA Data Challenge (LDC) has emerged as a crucial platform \cite{LDC}. Building on the legacy of the Mock LISA Data Challenges (MLDC) \cite{arnaud2006mock, babak2008mock, babak2010mock}, the LDC provides a comprehensive suite of simulated data sets mimicking real-world LISA observations. These data sets serve as a testing ground for researchers to develop and refine their data analysis methods.

The LDC is structured into various challenges, each designed to explore a specific aspect of LISA data analysis. LDC1, the first set of challenges, presents researchers with six distinct scenarios, each focusing on a single type of GW signal. For instance, LDC1-1 contains a single massive black hole binary (MBHB) signal along with simulated instrument noise. LDC2a presents a more complex challenge, tasking researchers with a data set containing a combination of 15 MBHBs, 30 million GBs, and instrument noise where the signal extraction and parameter estimation have to be a global fit of all parameters.

Several methods have been proposed for extracting GB signals, including maximum likelihood estimate (MLE) and Bayesian approaches: MLE methods are used to find the best matching simulated signal to the data  \cite{ bouffanais2016DE, zhang2021resolving, gao2023fast}, while Bayesian methods provide a posterior distribution that describes the uncertainty of the source parameters. Successful approaches to Bayesian analysis are Markov chain Monte Carlo (MCMC) based methods. Examples for MCMC based algorithms are blocked annealed Metropolis-Hastings (BAM) \cite{PhysRevD.75.043008, crowder2007genetic, littenberg2011detection}, an MCMC algorithm with simulated annealing, or the reversible jump Markov chain Monte Carlo (RJMCMC) \cite{10.1093/biomet/82.4.711, gamerman2006markov,littenberg2020global, littenberg2023prototype, karnesis2023eryn, tong2024transdimensional} method, which allows for varying parameter dimensions and thus variable numbers of GBs to construct the posterior distribution. Recently, a hybrid method has been introduced that leverages both MLE and MCMC \cite{strub2022, strub2023accelerating}. This approach is used to first find the MLE and then to compute the posterior distribution around the identified MLE with a parallelized MCMC method.

Methods to analyze the MBHB signals have historically used both MCMC-based or MLE-based algorithms, as detailed in various studies \cite{vecchio2004lisa, berti2005estimating, arun2006parameter, cornish2006mcmc,rover2007inference, feroz2009use, gair2009swarms, petiteau2010search, porter2015fisher}. In recent years, the focus has shifted from MLE algorithms towards MCMC algorithms, offering robust posterior distribution computations \cite{cornish2020black, marsat2021exploring, katz2020gpu, cornish2022low, katz2022fully}. The analysis with a more accurate LISA response including higher harmonic modes was recently explored in a noiseless data set \cite{katz2022fully}. Additionally, the impact of data gaps on the analysis of MBHBs, specifically in scenarios without noise, has been investigated \cite{dey2021effect}.

Recently, a prototype global analysis of MBHBs and GBs has been published \cite{littenberg2023prototype}. The pipeline is a blocked MCMC sampler to obtain a simultaneous global fit of the LDC2a data set. The method to extract the MBHBs is a parallel tempering MCMC algorithm \cite{cornish2022low}, and the GB extraction is based on a RJMCMC algorithm \cite{littenberg2020global}.

In this article, we present a novel pipeline to analyze together MBHB and GBs. The performances of the pipeline are evaluated with the LDC2a data set. Additionally, we propose a novel pipeline for extracting MBHBs within LISA data. We revisit MLE approaches by introducing a robust genetic algorithm-driven pipeline. This pipeline effectively extracts MBHBs from noisy data, incorporating scenarios with and without higher modes. The global fit pipeline combines the newly developed MBHB extraction method with the established GB algorithm from our previous work \cite{strub2023accelerating} and a generalized noise estimate algorithm.

In Sec. \ref{sec:bayes} we introduce the likelihood function, while in Sec. \ref{sec:global fit sec} we present the MBHB algorithm, changes made on the previous GBs algorithm, and describe the global fit pipeline. Our results, presented in Section \ref{sec:results}, include testing the MBHB algorithm on LDC1-1 and a self-injected higher mode signal, along with the global fit pipeline on the LDC2a data set. Additionally, Section \ref{sec:robustness} showcases the results of robustness tests of the pipeline. Finally, in Section \ref{sec:conclusion}, we engage in a discussion regarding the performance and future extensions of the pipeline.

\section{The Likelihood function}
\label{sec:bayes}
LISA data $d(t)$ will consist of the sum of multiple GW signals $s(t, \theta)$ and noise $n$ which can be written as $d(t) = \sum s(t,\theta) + n(t)$. For simplicity, we shall henceforth forego the time dependency notation, although $d$, $s(\theta)$, and $n$ remain time series.

The likelihood function $p(d | \theta)$ serves to quantify the probability of observing the data stream $d$ given the source parameters $\theta$ characterizing the GW signal. Employing the logarithm of the likelihood, denoted as $\log p(d | \theta)$, is prevalent owing to its computational tractability and is expressed as:

\begin{equation}
\label{eq:log-likelihood}
\log p(d | \theta) = -\frac{1}{2} \langle d-s(\theta) | d-s(\theta) \rangle,
\end{equation}

where $\langle x(t) | y(t) \rangle$ signifies the scalar product between two time domain signals $x(t)$ and $y(t)$. It is defined as:

\begin{equation}
\label{eq:scalar}
\langle x(t) | y(t) \rangle = 4 \mathcal{R} \left( \int_0^\infty \frac{\tilde{x}(f) \tilde{y}^*(f)}{S(f)} \, df \right).
\end{equation}

Here, $\tilde{x}(f)$ denotes the Fourier transform of $x(t)$, and $S(f)$ represents the one-sided power spectral density of the noise. The noise estimation is continuously refined during the analysis process, with the specifics regarding noise estimation discussed in Sec.\ref{sec:noise}.

To mitigate laser noise inherent in the laser measurements of LISA's arms, the technique of time-delay-interferometry (TDI) is employed, converting the six Doppler measurements into three observables: X, Y, and Z \cite{tinto1999cancellation, Armstrong_1999, estabrook2000time, dhurandhar2002algebraic, tinto2014time}. Consequently, the data $d$ and the signal $s(\theta)$ entail TDI responses across multiple channels. The scalar product here reads as:

\begin{equation}
\langle  d-s\left(\theta \right) |  d-s\left(\theta \right) \rangle = \sum_{\alpha \in \mathcal{M}} \langle  d_\alpha-s_\alpha\left(\theta \right) |  d_\alpha-s_\alpha\left(\theta \right) \rangle 
\end{equation}

where $\mathcal{M} = { X, Y, Z }$ represents the default TDI combination. Alternatively, $\mathcal{M} = { A, E, T }$ where:

\begin{equation}
    \begin{aligned}
 A &= \frac{1}{\sqrt{2}} \left( Z - X \right) \\
 E &= \frac{1}{\sqrt{6}} \left( X - 2Y + Z \right) \\
 T &= \frac{1}{\sqrt{3}} \left( X + Y + Z \right)
    \end{aligned}
 \end{equation}

These channels are specifically chosen for their lack of correlation with instrument noise assuming equal arm lengths and stationary noise \cite{prince2002lisa, vallisneri2005synthetic}. However, when transitioning to the analysis of actual observational data, these assumptions will not hold. Accounting for more realistic noise correlations among TDI data channels will increase the computational cost for likelihood evaluations. While we utilize $A$, $E$, and $T$, we only consider $A$ and $E$ for signals with frequencies $f < f_\ast / 2 = 1/(4 \pi L) \approx \SI{9.55}{mHz}$, given the suppression of the GW response for $T$ where $f_\ast$ is the transfer frequency and $L$ is the distance between two spacecraft of LISA \cite{littenberg2020global}.


\section{Time evolving global fit}
\label{sec:global fit sec}
In our pipeline to analyze data containing signals of multiple source types, we first discuss the noise estimation and the signal extraction methods individually. In the last subsection \ref{sec:global fit}, we present the global fit pipeline combining the individual algorithms.

\subsection{Noise Estimation}
\label{sec:noise}

The noise estimation process is performed across the full frequency band and undergoes a modified version as outlined in \cite{strub2023accelerating}. Here, we provide a concise overview of the pipeline along the modifications.

Initially, if any GW signals have been previously detected, we subtract them from the data. Specifically, in our scenario, we subtract the recovered GBs denoted as $\tilde\theta_{\text{GB,recovered}}$ and recovered MBHBs denoted as $\tilde\theta_{\text{MBHB,recovered}}$ from the data, and obtain the residual

\begin{widetext}
\begin{equation}
\begin{split}
    d_\text{residual} &= d - \sum\limits_{\theta \in \tilde\theta_{\text{GB,recovered}}} s_\text{GB}(\theta) - \sum\limits_{\theta \in \tilde\theta_{\text{MBHB,recovered}}} s_\text{MBHB}(\theta) = d - d_\text{GB} - d_\text{MBHB}
\end{split}
\end{equation}
\end{widetext}

Here, $s_\text{GB}(\theta)$ and $s_\text{MBHB}(\theta)$ represent the signal corresponding to the MLE of each recovered parameter set $\theta$ for GBs and MBHBs. The sum of all GB and MBHB signals are represented by $d_\text{GB}$ and $d_\text{MBHB}$ respectively.

Subsequently, the residual data $d_\text{residual}$ is utilized to estimate a noise curve across the frequency domain employing Welch's method. To address outlier peaks, particularly those associated with unresolved signals, a smoothing procedure is implemented. A frequency window comprising 30 bins is defined, and any values exceeding double the window's median are adjusted to the median value. This process is reiterated by shifting the window by 15 frequency bins until the entire Power Spectral Density (PSD) is smoothed. This PSD smoothing procedure is similar to conducting a running median as discussed in \cite{nissanke2012gravitational, karnesis2021characterization}.

Further smoothing is conducted using the Savitzky-Golay filter. Spline interpolation is then applied to derive a PSD estimate, leading to the residual noise curve.

Different from the algorithm described in \cite{strub2023accelerating}, while implementing Welch's method, we transition from a fixed number of 500 windows to a fixed number of $15\, 000$ samples per window. This change aims to generalize the noise estimate for varying observation times, considering that the number of frequency bins changes with $T_\text{obs}$. Consequently, the window length for the Savitzky-Golay filter is fixed at 10 frequency bins for frequencies below $\SI{0.8}{mHz}$ and at 70 bins for frequencies above, facilitating a consistent approach across different observation duration $T_\text{obs}$.

\begin{figure}[!ht]
\includegraphics[width=0.5\textwidth]{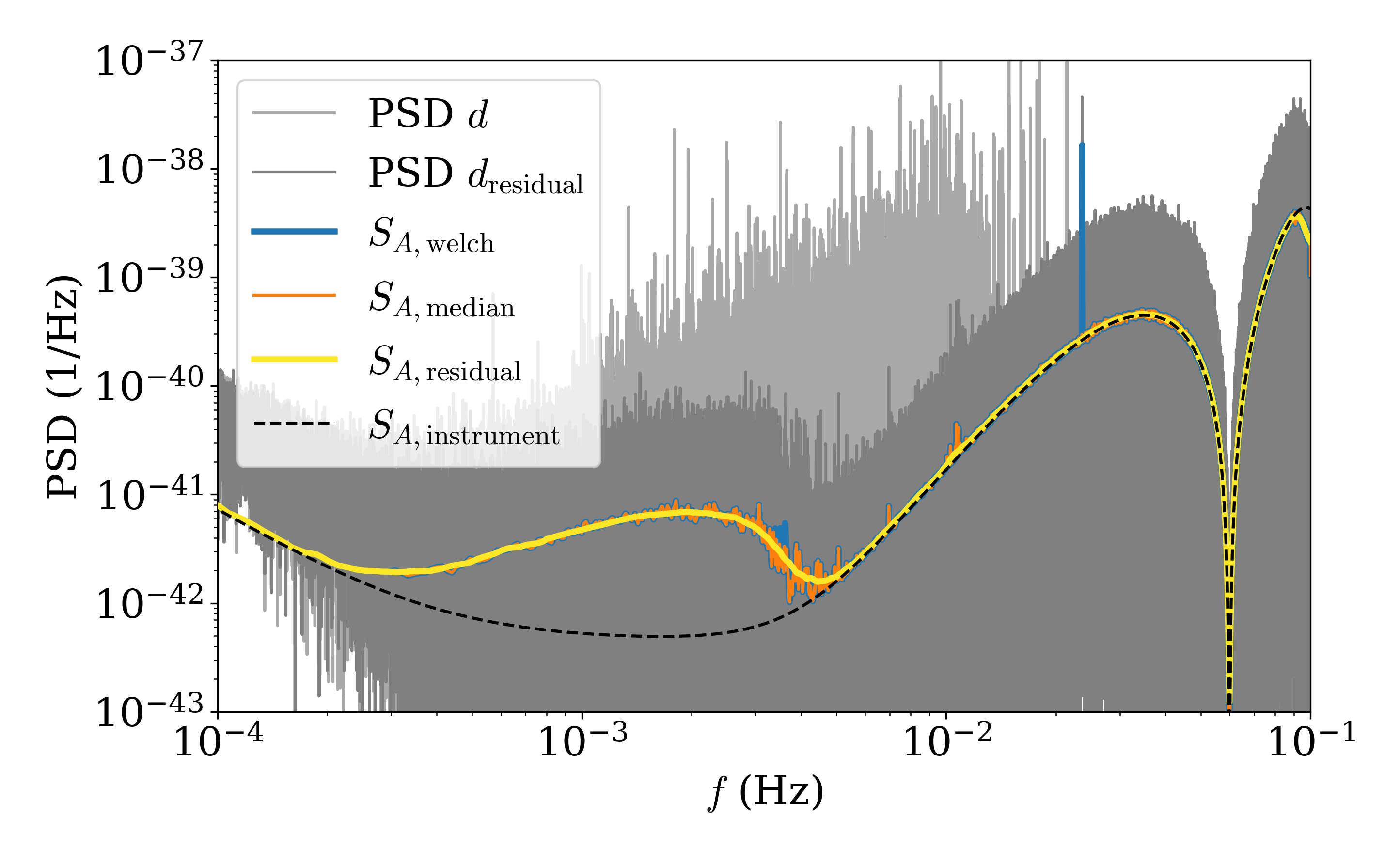}
\caption{The noise estimates and PSD of the TDI A channel within the $\SI{1}{yr}$ LDC2a data set are shown. $S_{A \text{,instrument}}$ represents the noise PSD utilized in generating the data set. The discrepancy observed between $S_{A \text{,residual}}$ and the instrument PSD within the $\SI{0.2}{mHz}$ to $\SI{5}{mHz}$ range is attributed to unresolved Galactic Backgrounds (GBs). It is anticipated that the majority of GBs within this frequency span remain unresolved, leading them to blend into the Galactic foreground noise. The blue spike is an unsolvable GB signal at $\SI{26}{mHz}$ due to an issue with the data set instead of the analysis. At \cite{strub_noise_visualizations} we provide videos and figures of the noise estimate for each week.}\label{fig:noise}
\end{figure}

In Fig. \ref{fig:noise} we present the noise estimate for the LDC2a data set. The noise estimates generally exhibit alignment with instrument noise, except for the $\SI{0.2}{mHz}$ to $\SI{4}{mHz}$ range, where deviations occur due to unresolved background signals. Similar computation procedures are employed for noise estimation in other TDI channels ($E$ and $T$).

\subsection{Extracting Massive Black Hole Binary signals}
\label{sec:mbhb}

To simulate the LISA response of a GW emitted by a MBHB we use the $\texttt{BBHx}$ \cite{katz2020gpu, katz2022fully} algorithm, accessible at \cite{michael_katz_2021}. This simulator utilizes 11 parameters $\theta_\text{MBHB} = \left\{\beta, \lambda, M, q, a_1, a_2, D_L, \iota, \phi_0, \psi, t_\text{ref} \right\}$ to model the GW signal. Here, $\lambda$ and $\beta$ correspond to the sky coordinates in terms of ecliptic longitude and ecliptic latitude, respectively. The total mass $M = m_1 + m_2$ represents the sum of the two masses of the binary, and $q$ denotes the mass ratio $q = m_2/m_1$. The spins $a_1$ and $a_2$ pertain to the objects with masses $m_1$ and $m_2$, respectively, and are aligned with the orbital angular momentum of the system; a negative spin indicates anti-parallel alignment. $D_L$ denotes the luminosity distance, $\iota$ represents the inclination angle, $\phi_0$ signifies the initial phase, and $\psi$ corresponds to the polarization angle. The parameter set $\theta_\text{MBHB}$ is defined within the prior $\Theta_\text{MBHB}$ as shown in Table \ref{tab:prior mbhb}. The prior is linearly uniform except for $\beta$, $\iota$, and $M$, which are sine-, cosine-, and log-uniform, respectively.

\begin{table}[!ht]
\caption{Boundaries of the prior distribution $\Theta_\text{MBHB}$ for MBHBs.}
\begin{ruledtabular}
\begin{tabular}{ccc}
           Parameter &  Lower Bound & Upper Bound  \\ \hline
    $\sin \beta$  &     $-1$   &  1\\
  $\lambda$&      $0$ & $2 \pi$\\
  $\log M$ ($\textup{M}_\odot$) & $\log 10^5$   &  $\log 10^8$\\
  $q$ & $1$   &  $10$\\
  $a_1$ & $-1$   &  $1$\\
  $a_2$ & $-1$   &  $1$\\
  $D_L$ (Gpc) & 0.5   &  1000 \\
         $\cos \iota$&      $-1$  & 1\\
        $\phi_0$&          $-\pi$  & $\pi$\\
         $\psi$&          0  & $\pi$ \\
  $t_\text{ref}$ & $\min (t_\text{segment})$  &  $\max(t_\text{segment}) $\\
\end{tabular}
\label{tab:prior mbhb}
\end{ruledtabular}
\end{table}

To obtain the MLE we can maximize the signal-to-noise ratio (S/N) defined as
\begin{equation}
\label{eq:SNR}
\rho =  \frac{\langle  d |  s\left(\theta^\prime \right) \rangle}{\sqrt{\langle  s\left(\theta^\prime \right) |  s\left(\theta^\prime \right) \rangle}} =  \frac{\langle  d |  s\left(\theta \right) \rangle}{\sqrt{\langle  s\left(\theta \right) |  s\left(\theta \right) \rangle}}.
\end{equation}
which is independent of $D_L$ with $\theta^\prime = \theta \setminus \{ D_L\}$ and obtain analytically the luminosity distance which maximizes the likelihood with 

\begin{equation}
\label{eq:maxD}
D_L =  \frac{\langle  s\left(\theta^\prime \right) |  s\left(\theta^\prime \right) \rangle }{\langle  d |  s\left(\theta^\prime \right) \rangle}.
\end{equation}

The MBHB signal extraction uses a time-evolving algorithm that extracts MBHBs within LISA data time segments. This approach reflects a more realistic scenario where the inspiral phase of MBHBs may be present within the data but remain unextracted since the merger has not been recorded yet.

In Algorithm \ref{alg:MBHB local}, we provide pseudo-code for the search algorithm, using differential evolution (DE) \cite{storn1997differential} to maximize eq. \ref{eq:SNR}, aiming to obtain the MLE of MBHBs within a time series segment $d_\text{segment}$ spanning a set time $t_\text{segment}$ of the data. To mitigate the potential issue of a merger occurring at the border of $d_\text{segment}$, we extend the segment to analyze $d_\text{analyze}$, incorporating padded regions before and after the segment. Both the segment and padding lengths are adjustable, but for this work, we use a one-week segment with one week of padding before and one day of padding after the segment, resulting in a total analysis window length of 15 days. The padding before the merger can be extended as desired until the start of measurement since MBHBs that merged before $d_\text{segment}$ have already been extracted. Conversely, the padding duration after $d_\text{segment}$, being in the future, is limited by available data. We deem a padding duration of $\SI{1}{day}$ after $d_\text{segment}$ to be sufficient.

\begin{algorithm}[!ht]
\caption{The MBHB search algorithm to analyze a data segment $d_\text{analyze}$.}\label{alg:MBHB local}
$\textbf{Function} \ \textbf{\textit{MBHB\_search}}(d_\text{analyze})$\\
$\tilde\theta_\textrm{found} \gets \{\,\}$\\
$d_\textrm{residual} \gets d_\textrm{analyze}$\\
$\rho_\textrm{max} \gets \rho_\textrm{threshold} + 1$\\
\textbf{while} $\rho_\textrm{max} > \rho_\textrm{threshold}$  \textbf{do} \\
    \hspace*{1em} $\tilde\theta^\prime_\textrm{MLEs} \gets \{\,\}$\\

    \hspace*{1em}  \textbf{for $j$ in} $\{1,2,..., n_\textrm{searches}\}$  \textbf{do} \\
        \hspace*{2em} $\theta^\prime_{\textrm{init}} $ randomly drawn from prior\\
        \hspace*{2em} $\theta^\prime_\textrm{MLE} \gets  \argmax\limits_{\theta^\prime}  \rho (\theta^\prime, d_\textrm{residual}) $ using DE with $\theta^\prime_{\textrm{init}}$\\
        \hspace*{2em} \textbf{if} $\rho (\theta^\prime_\textrm{MLE}, d_\textrm{analyze}) \leq \rho_\textrm{threshold} \textbf{ and } j> 1 \textbf{ do}$\\
            \hspace*{3em} $\textbf{ return } \tilde\theta_\textrm{found}$\\
        \hspace*{2em} $\tilde\theta^\prime_\textrm{MLEs} \gets \tilde\theta^\prime_\textrm{MLEs} \cup \{\theta^\prime_\textrm{MLE}\}$\\
    \hspace*{1em} \textbf{end for}\\

    \hspace*{1em} $\theta^\prime_\textrm{MLE} \gets \argmax\limits_{\theta^\prime \in \tilde\theta^\prime_\textrm{MLEs}} \rho (\theta^\prime, d_\textrm{residual})$\\ 
    \hspace*{1em} $\rho_{max} \gets \rho (\theta^\prime_\textrm{MLE}, d_\textrm{analyze})$\\
    \hspace*{1em} \textbf{if} $\rho_{max} \leq \rho_\textrm{threshold} \textbf{ do}$\\
        \hspace*{2em}$\textbf{return } \tilde\theta_\textrm{found}$\\
    \hspace*{1em} Compute $D_L$ according to eq. \eqref{eq:maxD} with $\theta^\prime_\textrm{MLE}$\\
    \hspace*{1em} $\theta_{\textrm{MLE}} \gets \theta^\prime_{\textrm{MLE}} \cup \{D_L\}$\\
    
    \hspace*{1em} $\tilde\theta_\textrm{found} \gets \tilde\theta_\textrm{found} \cup \{\theta_{\textrm{MLE}}\}$\\
    \hspace*{1em} $d_\textrm{residual} \gets d_\textrm{residual} - s(\theta_\textrm{MLE})$\\
\textbf{end while}\\
$\textbf{return } \tilde\theta_\textrm{found}$
\end{algorithm}

To obtain a good fit with a high likelihood we set the number of searches for each signal to $n_\textrm{searches} = 5$. This repeats each search 5 times with different random initial parameters $\theta^\prime_{\textrm{init}}$ and then selects the result with the highest likelihood.

The MLE can then be used to subtract the MBHBs from the data containing signals of multiple source types to analyze the other signals like GBs without the presence of the MBHBs. Furthermore, the MLE can be used to obtain an uncertainty estimate through the Fisher Information Matrix (FIM) or any MCMC algorithm with the MLE as a proposal to improve the sampling efficiency.

The uncertainty estimate with FIM is generally not valid by wrongfully assuming Gaussian distribution of errors \cite{porter2015fisher}. Furthermore, we neglect correlations between sources. For completeness, we still use the FIM to provide a fast and simple uncertainty estimate even though an uncertainty estimate is not necessarily needed for the global fit. We compute the uncertainties by taking the square root of the inverse of the FIM, defined as:

\begin{equation}
F_{ij} = \langle \partial_i s(\theta) | \partial_j s(\theta) \rangle,
\end{equation}

where $\partial_i$ represents the partial derivative concerning the $i$\textsuperscript{th} component of the parameter vector $\theta$. The derivatives of the FIM are computed using the second-order center finite difference method with a step size of $10^{-9}$ times the exploration space determined by the prior distribution $\Theta_{\text{MBHB}}$. The estimated uncertainty vector is denoted as $\sigma = \sqrt{\text{diag}(F^{-1})}$.

\subsection{Extracting Galactic Binary signals}
\label{sec:full galaxy}

The GB extraction algorithm, to construct $d_\text{GB}$, is described in an earlier paper \cite{strub2023accelerating}. The algorithm splits the frequency domain data into multiple segments $B_\text{search}$ to be analyzed. Where one segment is double the size of the maximum expected bandwidth of a signal within the segment. The segments are split into two groups of non-neighboring segments called $B_\text{even}$ and $B_\text{odd}$. All even segments are first analyzed in parallel and then the $B_\text{odd}$ segments are analyzed with the signals found within the even segments subtracted. Lastly, the even segments are reanalyzed with the signals found within the odd segments subtracted.

Simulating a GB signal involves eight parameters $\theta_\text{GB} = \left\{\mathcal{A}, \lambda, \beta, f, \dot{f}, \iota, \phi_0, \psi\right\}$ \cite{cornish2007tests}. Here $\mathcal{A}$ represents the amplitude, $\lambda$ and $\beta$ correspond to the sky coordinates in terms of ecliptic longitude and ecliptic latitude, respectively. The parameter $f$ signifies the frequency of the GW, while $\dot{f}$ denotes the first-order frequency derivative. The inclination angle is denoted by $\iota$, $\phi_0$ signifies the initial phase and $\psi$ corresponds to the polarization angle. The prior distribution is listed in Table \ref{tab:prior}.

\begin{table}[!ht]
\caption{Boundaries of the prior distribution $\Theta_\text{GB}$.}
\begin{ruledtabular}
\begin{tabular}{ccc}
           Parameter &  Lower Bound & Upper Bound  \\ \hline
    $\sin \beta$  &     $-1$   &  1\\
  $\lambda$&      $-\pi$ & $\pi$\\
  $f$ & $\SI{0.3}{mHz}$  &  $f_\text{Nyquist}$\\
  $\dot{f}$ &  $ -5 \cdot 10^{-6} f^{13/3}$  & $1.02 \cdot 10^{-6} f^{11/3}$ \\
  $\log \mathcal{A}$ & $\log \mathcal{A} (\rho = 7)$ & $\log \mathcal{A} (\rho = 1000)$ \\
         $\cos \iota$&      $-1$  & 1\\
        $\phi_0$&          0  & 2$\pi$\\
         $\psi$&          0  & $\pi$ \\
\end{tabular}
\label{tab:prior}
\end{ruledtabular}
\end{table}

The amplitude boundary is determined by a lower and upper bound signal to noise ratio $\rho$ and is related to the amplitude by \cite{littenberg2020global}

\begin{equation}
\mathcal{A} \left( \rho \right) = 2 \rho \left( \frac{S\left(f\right)} {T_{obs} \, \sin^2\left(f / f_\ast \right)} \right)^{1/2}.
\end{equation}

\subsection{Pipeline of the time evolving global fit}
\label{sec:global fit}

Combining individual algorithms to analyze data sets containing signals of multiple source types, we introduce a time-evolving pipeline that dynamically updates signal extraction as new data arrives.

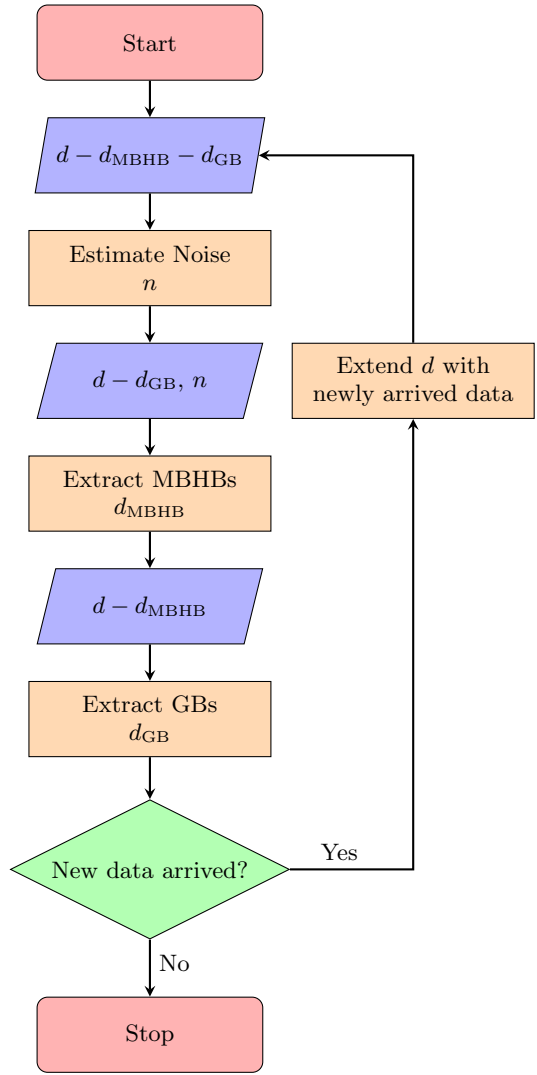
\begin{figure}[ht!]
\caption{\label{fig:flow} Flow chart of the time-evolving extraction of MBHBs and GBs and noise estimate. The parallelograms represent the inputs for the processes, which are represented by rectangles. For the first iteration $d_{\mathrm{MBHB}} = d_{\mathrm{GB}} = 0$ which are then updated by the pipeline.}
\begin{tikzpicture}[node distance=1.5cm]

\node (start) [startstop, yshift=-0.5cm] {Start};
\node (in1) [io, below of=start] {$d-d_{\mathrm{MBHB}}-d_{\mathrm{GB}}$};
\node (pro1) [process, below of=in1] {Estimate Noise  \\ $n$};
\node (in2) [io, below of=pro1] {$d-d_{\mathrm{GB}}$, $n$};
\node (pro2) [process, below of=in2] {Extract MBHBs  \\ $d_{\mathrm{MBHB}}$};
\node (in3) [io, below of=pro2] {$d-d_{\mathrm{MBHB}}$};
\node (pro3) [process, below of=in3] {Extract GBs  \\ $d_{\mathrm{GB}}$};
\node (pro4) [process, right of=in2, xshift=2cm] {Extend $d$ with newly arrived data};
\node (if) [decision,aspect=2, below of=pro3, yshift=-0.5cm] {New data arrived?};
\node (stop) [startstop, below of=if, yshift=-0.7cm] {Stop};

\draw [arrow] (start) -- (in1);
\draw [arrow] (in1) -- (pro1);
\draw [arrow] (pro1) -- (in2);
\draw [arrow] (in2) -- (pro2);
\draw [arrow] (pro2) -- (in3);
\draw [arrow] (in3) -- (pro3);
\draw [arrow] (pro3) -- (if);
\draw [arrow] (if) -- node[pos=0.4,right] {No} (stop);
\draw [arrow] (if) -| node[pos=0.2,above] {Yes} (pro4);
\draw [arrow] (pro4) |-  (in1);

\end{tikzpicture}
\end{figure}

In Fig. \ref{fig:flow} we illustrate the pipeline through a flow diagram. For a likelihood-based signal extraction we need a noise estimate. Initially, we estimate the noise using the original data since at the beginning no signals are extracted and therefore $d_{\mathrm{MBHB}} = d_{\mathrm{GB}} = 0$ for the first iteration.

After estimating the noise we first extract the MBHBs since a new MBHB merger could have been recorded, and these mergers are so loud that they would tamper significantly with the GB extraction. Moreover, since the GBs are quasi monochromatic, their signal does not change much from one time step to the next one. Therefore the GB extraction from the previous time set is already a good fit to the updated data and does not tamper much with the MBHB extraction.

In each iteration, the complete data set available up to that moment is reanalyzed. With the updated $d_\text{GB}$ incorporating an increased number of subtracted GB signals, there is reduced interference in the MBHB signal extraction process. Consequently, it is recommended to reassess all MBHBs as well to ensure optimal analysis outcomes.

Subsequently, we subtract the identified MBHBs $d_\text{MBHB}$ from the data and proceed to extract the GBs. The GB algorithm autonomously generates an on-the-fly noise estimate for each frequency segment, obviating the need for the noise estimate $n$ across the full frequency spectrum \cite{strub2023accelerating}.

Once new data arrived, we update the data set $d$ and go to the beginning of the pipeline as shown in Fig. \ref{fig:flow}.

\section{Results}
\label{sec:results}

Since the algorithms for the noise estimate and GBs have already been discussed in \cite{strub2022, strub2023accelerating} we now first show the results of the MBHB part individually. Later we show the results for the LDC2a containing overlapping GB and MBHB signals.

\subsection{MBHBs}
\label{sec:results mbhb}

\begin{table*}[!htbp]
\caption{The injected and recovered parameters of the MBHB of the LDC1-1 data challenge. The uncertainties are computed with the inverse of the Fisher Information Matrix.}
\begin{ruledtabular}
\begin{tabular*}{\textwidth}{@{\extracolsep{\fill}}cccc@{\extracolsep{\fill}}}
Parameter     & Injected & Recovered (2,2) mode & Recovered higher modes \\ \hline
$m_1$ ($\textup{M}_\odot$) $\times10^{-6}$    &  \num{2.599137} & \num{2.60 \pm 0.02} & \num{2.523 \pm 0.005} \\
$m_2$ ($\textup{M}_\odot$) $\times10^{-6}$    &  \num{1.242861} & \num{1.239 \pm 0.008} & \num{1.275 \pm 0.002} \\
$a_1$ $\times10^{1}$     &  \num{7.534822} & \num{7.6 \pm 0.2}  & \num{6.66 \pm 0.03} \\
$a_1$ $\times10^{1}$     &  \num{6.215875} & \num{6.0 \pm 0.5}  & \num{8.33 \pm 0.04} \\
$D_L$ (Gpc) $\times10^{-1}$    &  \num{5.600578} & \num{5 \pm 4}  & \num{3.5 \pm 0.4} \\
$\iota$     &  \num{1.224532} & \num{1.3 \pm 0.3}  & \num{1.35 \pm 0.03} \\
$\lambda$     &  \num{3.509100} & \num{3.49 \pm 0.07}  & \num{3.526 \pm 0.004} \\
$\beta$ $\times10^{1}$     &  \num{2.926963} & \num{2.7 \pm 0.6}  & \num{2.86 \pm 0.03} \\
$t_c$ (s) $\times10^{-7}$   &  \num{2.496000} & \num{2.495999 \pm 0.000003}  & \num{2.4960004 \pm 0.0000002} \\
\end{tabular*}
\label{tab:mbhb}
\end{ruledtabular}
\end{table*}


Further, we test the pipeline on the signal simulated considering higher modes $(l,m) = \{(2,2), (3,3), (4,4), (2,1), (3,2), (4,3)\}$ where the MBHB simulation of the LDC1-1 and LDC2a data sets only considers the primary $(l,m) = (2,2)$ mode \cite{katz2022fully}.

In the following, we present the results of the MBHB algorithm described in \ref{sec:mbhb} for the LDC1-1 data set, which consists of instrument noise and a single MBHB signal. As mentioned in \cite{katz2022assessing} the signal simulation also used in this work sets $t_\text{ref}$ at $f_\text{max}$ where $f_\text{max}$ is set internally which is different from the signal in LDC1-1 is simulated where $f^2 A_{22}$ is maximized instead. Therefore to avoid discrepancies due to using different simulators we subtract the MBHB signal from the LDC data and self-inject the signal with the same parameters using \texttt{BBHX} for the signal simulations.

\begin{figure}[!htbp]
\minipage{0.5\textwidth}
  \includegraphics[width=0.9\linewidth]{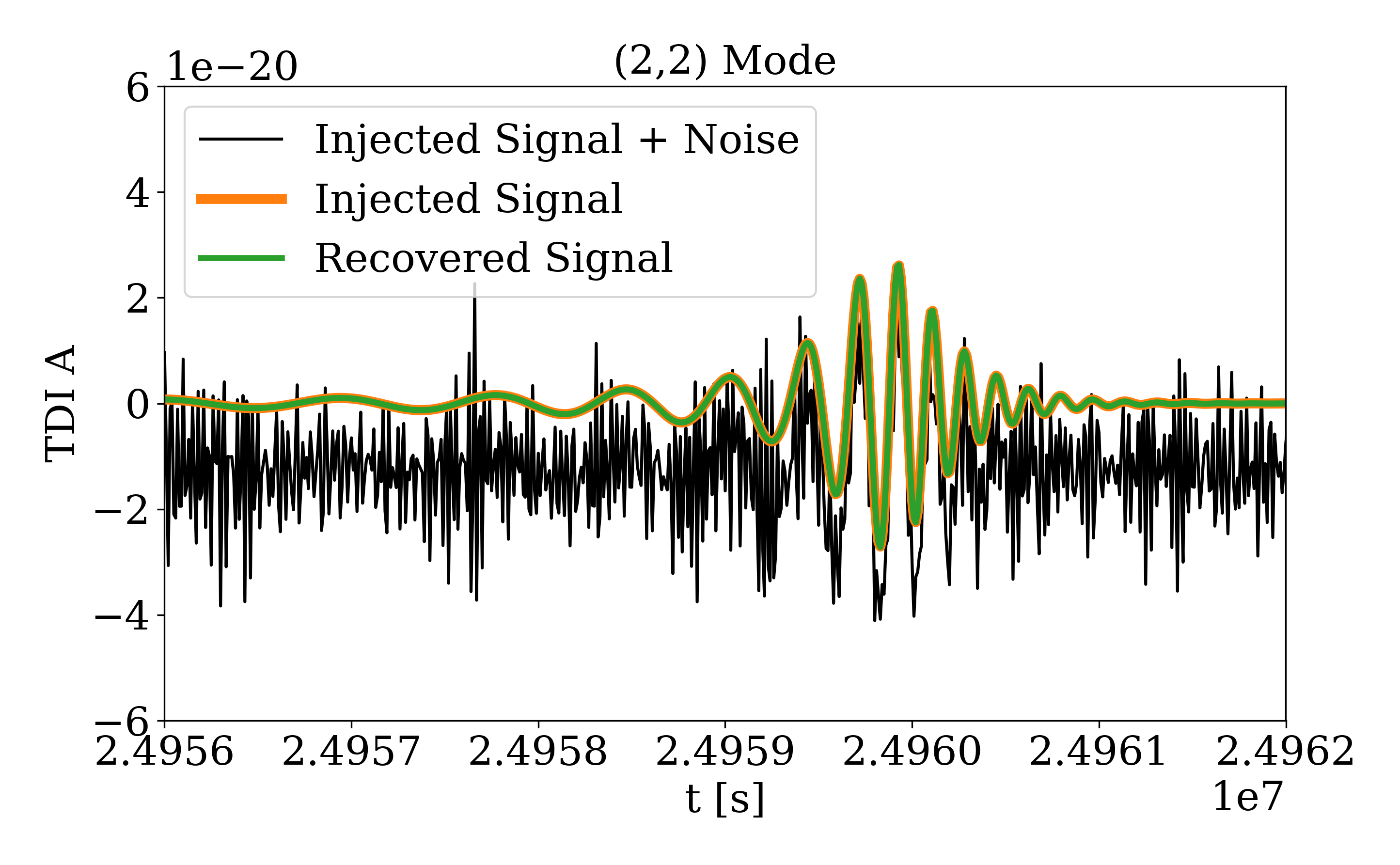}
\endminipage\vfill
\minipage{0.5\textwidth}
  \includegraphics[width=0.9\linewidth]{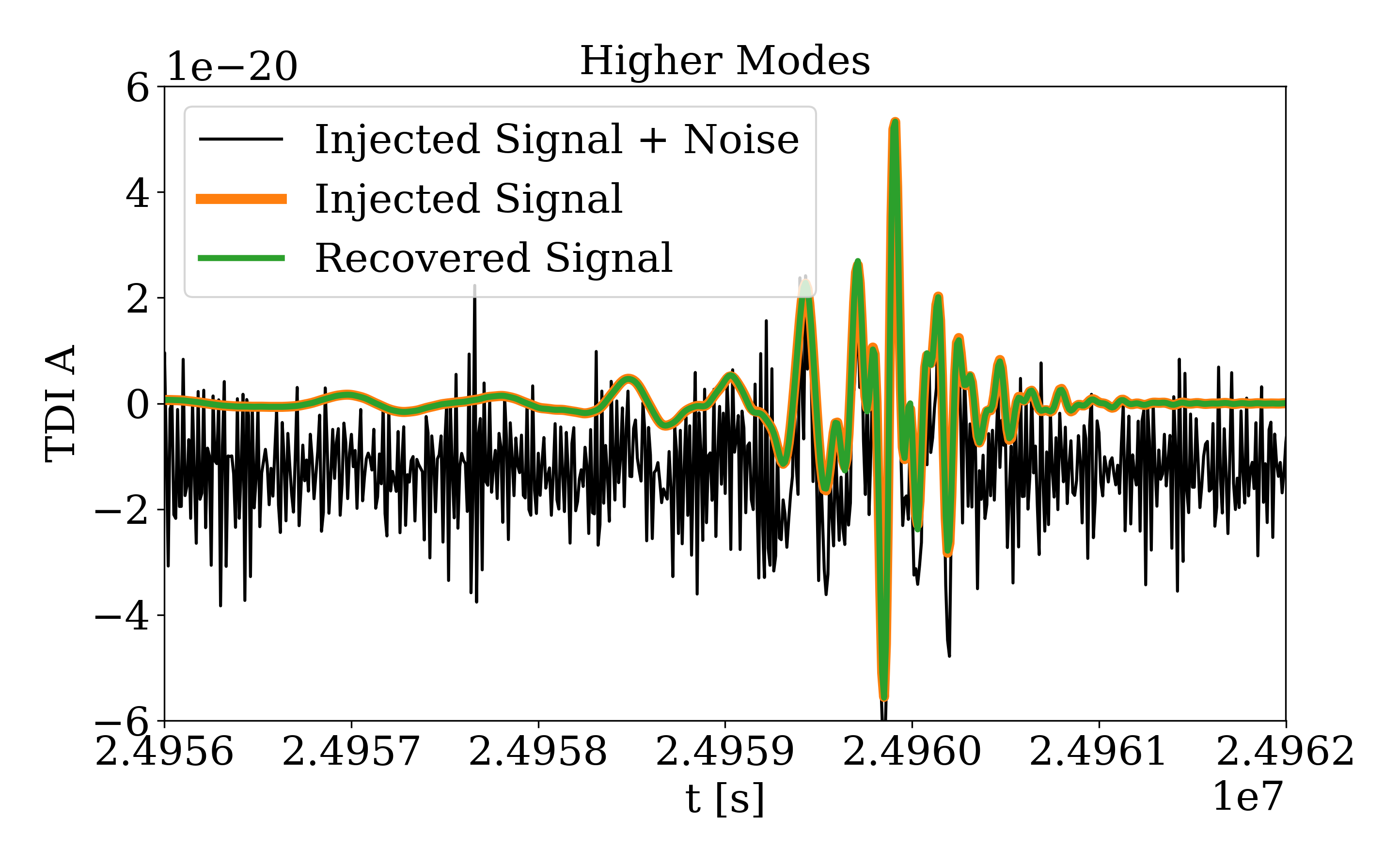}
\endminipage\vfill
\caption{LDC1-1 TDI A component at the merger of the MBHB signal. The black line is the data. The thicker orange line is the self injected signal according to the parameters given by the LDC catalog and the thin green line is the recovered signal.}
\label{fig:Radler TDI A}
\end{figure}

Since the observation time $T_\text{obs} = 1 yr$ is rather long we split the data into segments of $\SI{1}{week}$ duration.

In Fig. \ref{fig:Radler TDI A} we present the found signal alongside the injection and the data. All analyses have been conducted using noisy data. The padding is usually set to 7 days before and 1 day after the segment but for the analysis considering higher harmonic modes, we obtained better results by extending the padding before the segment to 28 days.

In Table \ref{tab:mbhb} we present the recovered parameters and their uncertainties. Since we compute the uncertainties by taking the inverse of the FIM we wrongfully assume a Gaussian posterior distribution for all parameters. Nonetheless, the obtained uncertainties are consistent with uncertainties computed with an MCMC sampler as presented in \cite{katz2022fully}.

\subsection{Multiple type signal sources}

\begin{figure*}[!htbp]
\includegraphics[width=1\textwidth]{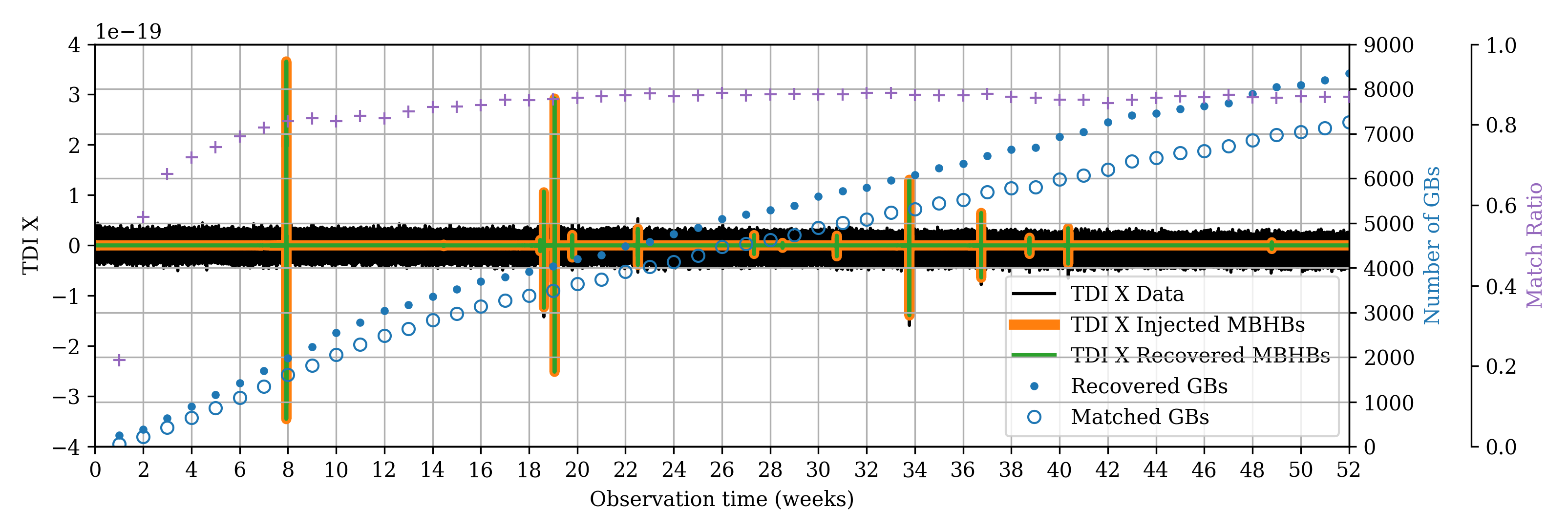}
\caption{The weekly results of the mixed signal LDC2a data set. In black we show the TDI A component of the LDC2a data set to be analyzed and in orange and green we show the injected and resolved MBHBs $d_\text{MBHB}$. The blue dots and circles are the recovered and matched GBs respectively. The purple crosses are the match ratio of matched GBs divided by recovered GBs.}
\label{fig:weekly_results} 
\end{figure*}

\begin{figure*}[!htbp]
\includegraphics[width=1\textwidth]{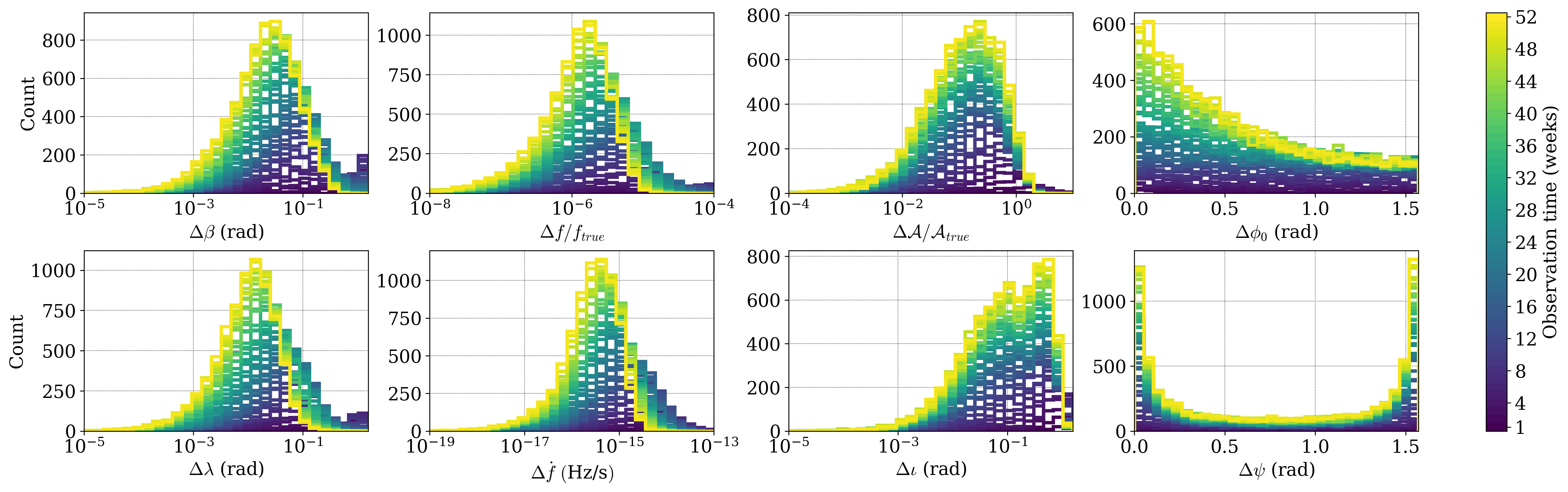}
\caption{Histogram of the difference of resolved and injected parameters of matched signals for different observation times. Note that for $f$ and $\mathcal{A}$ we present the relative difference.}
\label{fig:error_hist} 
\end{figure*}

Here we show the results of our analysis of the LDC2a containing overlapping GB and MBHB signals and instrument noise. We analyzed for GBs across the entire frequency spectrum $f \in [\SI{0.3}{mHz}, \SI{100}{mHz}]$. This range spans from regions where GB detection is improbable due to instrument noise to the Nyquist frequency. The extraction of MBHBs was carried out in time domain segments of the entire one-year data set.

Since the pipeline as described in section \ref{sec:global fit} is a time-evolving algorithm where the pipeline in Fig. \ref{fig:flow} runs once for each time step. We decide to analyze the data in weekly steps. Therefore we can set a realistic premise to analyze data as new data is coming in as in a realistic setup with LISA being online.

For both the MBHB and GB search algorithms we set $\rho_\text{threshold} = 9$ as the threshold to accept a found signal within the set of recovered signals. To simulate the LISA response of MBHBs we use the BBHX implementation \cite{michael_katz_2021} and for the GBs we use fastGB provided by the LDC \cite{LDC}.

\begin{figure*}[!htbp]
\includegraphics[width=1\textwidth]{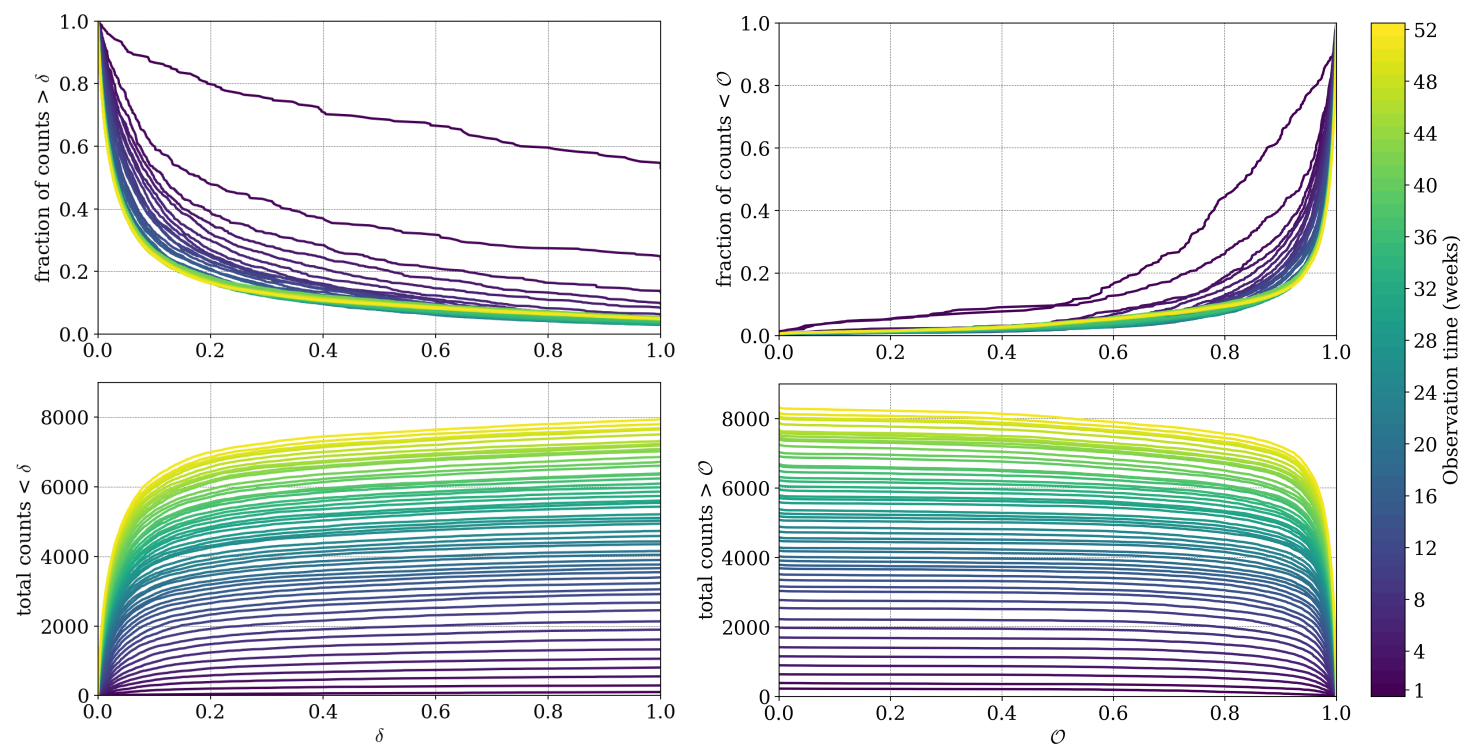}
\caption{\label{fig:CDF} The top plots illustrate the cumulative distribution function of scaled error $\delta$ and overlap $\mathcal{O}$ of all extracted GBs, while the bottom plots depict the survival function. Each color represents a different number of weeks of observation time.}
\end{figure*}

To analyze the accuracy of the recovered GB signals $\theta_\text{rec}$ we can match them with the injected parameter sets $\theta_{inj}$. This is only possible with simulated data. In other publications, the overlap $\mathcal{O}$ is used to define a metric of similarity between two signals

\begin{equation}
\mathcal{O}(s(\theta_\text{rec}),s(\theta_\text{inj})) = \frac{ \langle  s(\theta_\text{rec})), s(\theta_\text{inj}) \rangle} { \sqrt{\langle  s(\theta_\text{rec}), s(\theta_\text{rec}) \rangle \langle  s(\theta_\text{inj}), s(\theta_\text{inj}) \rangle}}.
\end{equation}

which is independent of the amplitude of the signals. Therefore in this work, we use the scaled error

\begin{equation}
\delta(s(\theta_\text{rec}),s(\theta_\text{inj})) = \frac{ \langle  s(\theta_\text{rec})-s(\theta_\text{inj}),  s(\theta_\text{rec})-s(\theta_\text{inj}) \rangle} { \sqrt{\langle  s(\theta_\text{rec}), s(\theta_\text{rec}) \rangle \langle  s(\theta_\text{inj}), s(\theta_\text{inj}) \rangle}}\label{eq:match_metric}
\end{equation}

which is dependent on the amplitude of the signals and does not run the risk of matching an injected parameter set with low amplitude which is part of the foreground noise. Particularly at lower frequencies, where there are fewer frequency bins and low amplitude signals are prevalent, there exists a heightened risk of substantial overlap with a recovered signal, despite the scaled error not indicating a match.

In Fig. \ref{fig:CDF} we observe a rapid change of the cumulative distribution function until $T_\text{obs} = \SI{10}{weeks}$ where the scaled error rapidly declines and for $T_\text{obs} > \SI{26}{weeks}$ onwards the cumulative distribution function stabilized. The steady increase in total counts with longer observation times is as expected. We define a recovered GB as matched if it forms a pair with an injected parameter set $\theta_\text{inj}$ where $\mathcal{O}(s(\theta_\text{rec}),s(\theta_\text{inj})) < 0.3$.

The results of the weekly analysis of the LDC2a data set are shown in Fig. \ref{fig:weekly_results}. As expected there is a steady increase in the amount of recovered signals as the observation time increases.

It is noteworthy that from week 16 onwards, the match ratio (ratio of matched signals to extracted signals) stabilizes within the range of $85\%$ and $88\%$. Before this, the match ratio is lower, indicating that the recovered signals do not correspond well to the injected ones. This observation is also evident in Fig. \ref{fig:error_hist}, where analyses with $T_\text{obs} < \SI{16}{weeks}$ exhibit a high number of signals with low accuracy in sky location. Also from week 16 onwards, the number of recovered GBs and matched GBs increases by roughly 129 and 115 for each week.

Even though there is a MBHB merging right at $t_c = \SI{19.06}{week}$, the analysis of the data set with an observation time of 19 weeks shows no change in the trend of the amount of recovered signals nor on the match rate. This shows that the pipeline successfully extracts GBs without interference from this particular MBHB merging soon afterwards. This is because the signal of this MBHB until $T_\text{obs} = \SI{19}{weeks}$ is only in the inspiral phase and its frequency is $\SI{0.4}{mHz}$ or less. This is well visible in Fig.\ref{fig:noise_weekly} which we will discuss later. Although there were no issues with loud MBHBs merging immediately after the observed time segment in this data set, one could consider extending the search to include MBHBs with merger times that fall outside of the observed time segment.
 
Also observable from Fig. \ref{fig:error_hist} is that for all parameters there is a clear trend of decreasing errors with longer observation times, as expected. The mostly flat histograms for $\phi_0$ and $\psi$ for $T_\text{obs} < \SI{10}{weeks}$ stem from their intrinsic degeneracy \cite{strub2023accelerating}. Nonetheless, as $T_\text{obs}$ increases, it becomes clear that this degeneracy progressively diminishes.

\begin{figure}[!htbp]
\includegraphics[width=0.5\textwidth]{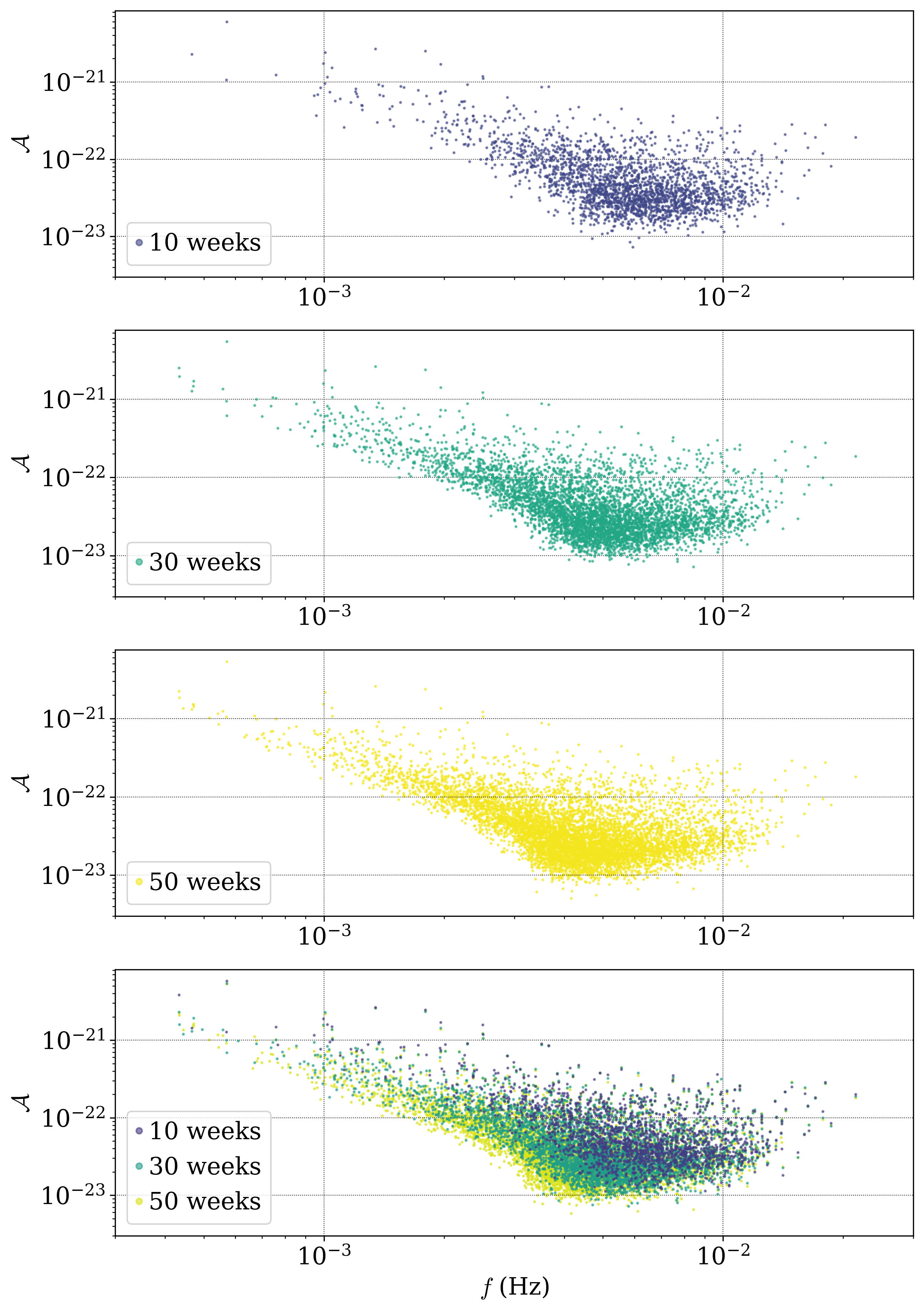}
\caption{Scatter plot of recovered GBs of the LDC2a data set of different observation times. Each of the shown recovered signals matched with an injected signal with $\delta < 0.3$. The bottom plot shows an overlap of the three analyses.}
\label{fig:amplitude_scatter}
\end{figure}

\begin{figure}[!htbp]
\includegraphics[width=0.5\textwidth]{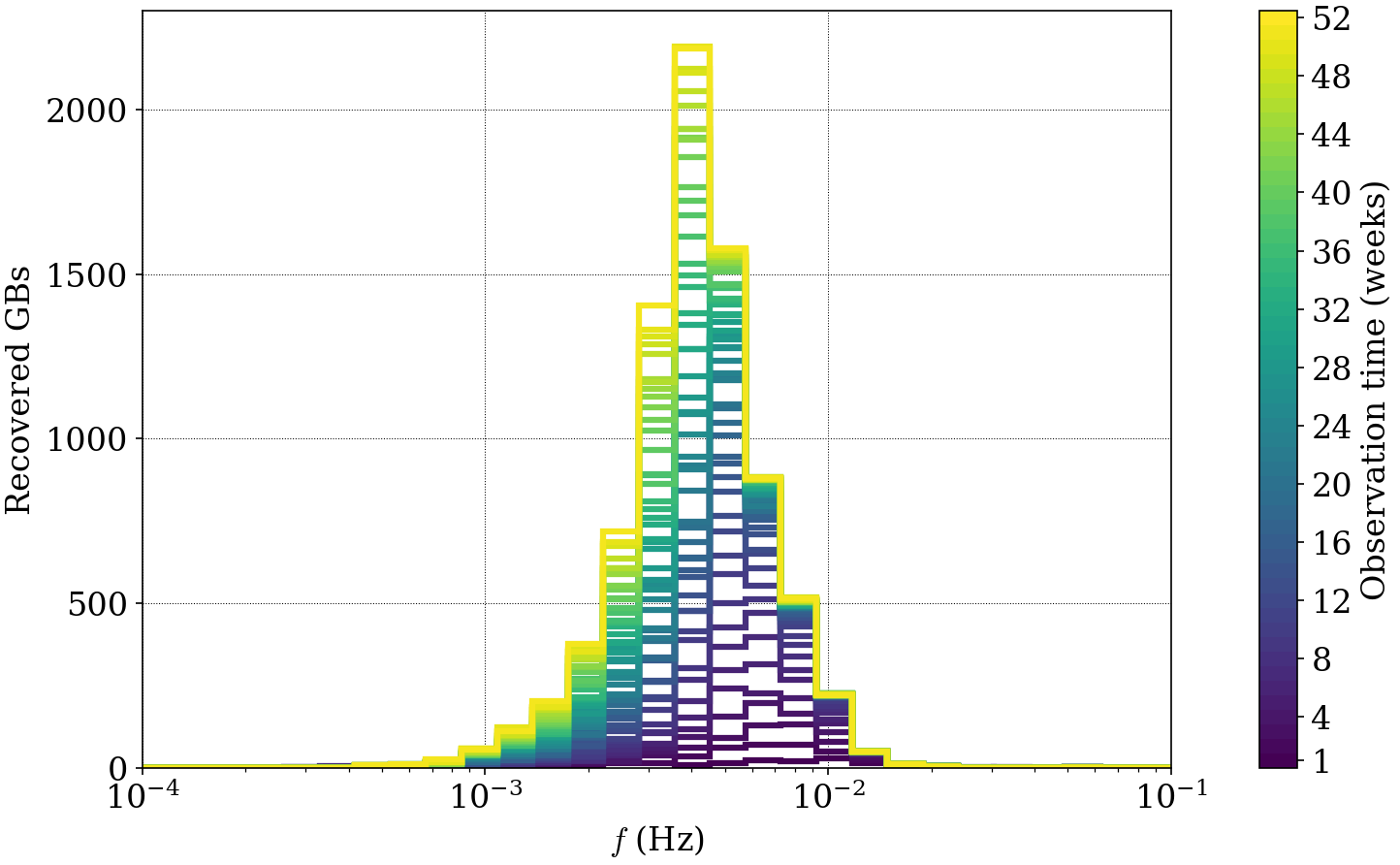}
\caption{Histogram of the recovered GBs of the LDC2a data set of all weekly analyses.}
\label{fig:frequency_histogram}
\end{figure}

In Fig. \ref{fig:amplitude_scatter} we present a scatter plot of $\mathcal{A}$ and $f$ for three different observation times. It is well noticeable that the number of recovered GBs increases overall and that also GBs of lower $\mathcal{A}$ are recovered. An increase of $T_\text{obs}$ increases the SNR which allows for the detection of fainter signals, as well as signals of lower frequency $f$. Furthermore, Figure \ref{fig:frequency_histogram} depicts a noticeable trend where the distribution of recovered GBs shifts toward lower frequencies as observation time lengthens. Specifically, at $T_\text{obs} = \SI{30}{weeks}$, the majority of resolved GBs are centered around a frequency of approximately $\SI{5}{mHz}$, whereas by $T_\text{obs} = \SI{52}{weeks}$, the frequency peak shifts to around $\SI{4}{mHz}$. This shift occurs as the number of recovered GBs converges towards the number of injected GBs with $f > \SI{5}{mHz}$, underscoring no further GBs at this or higher frequencies are left to be resolved. This pattern highlights the critical role of mission duration in the effective recovery of GBs, especially at frequencies around $\SI{4}{mHz}$.

\subsection{Noise}

\begin{figure}[!htbp]
\includegraphics[width=0.5\textwidth]{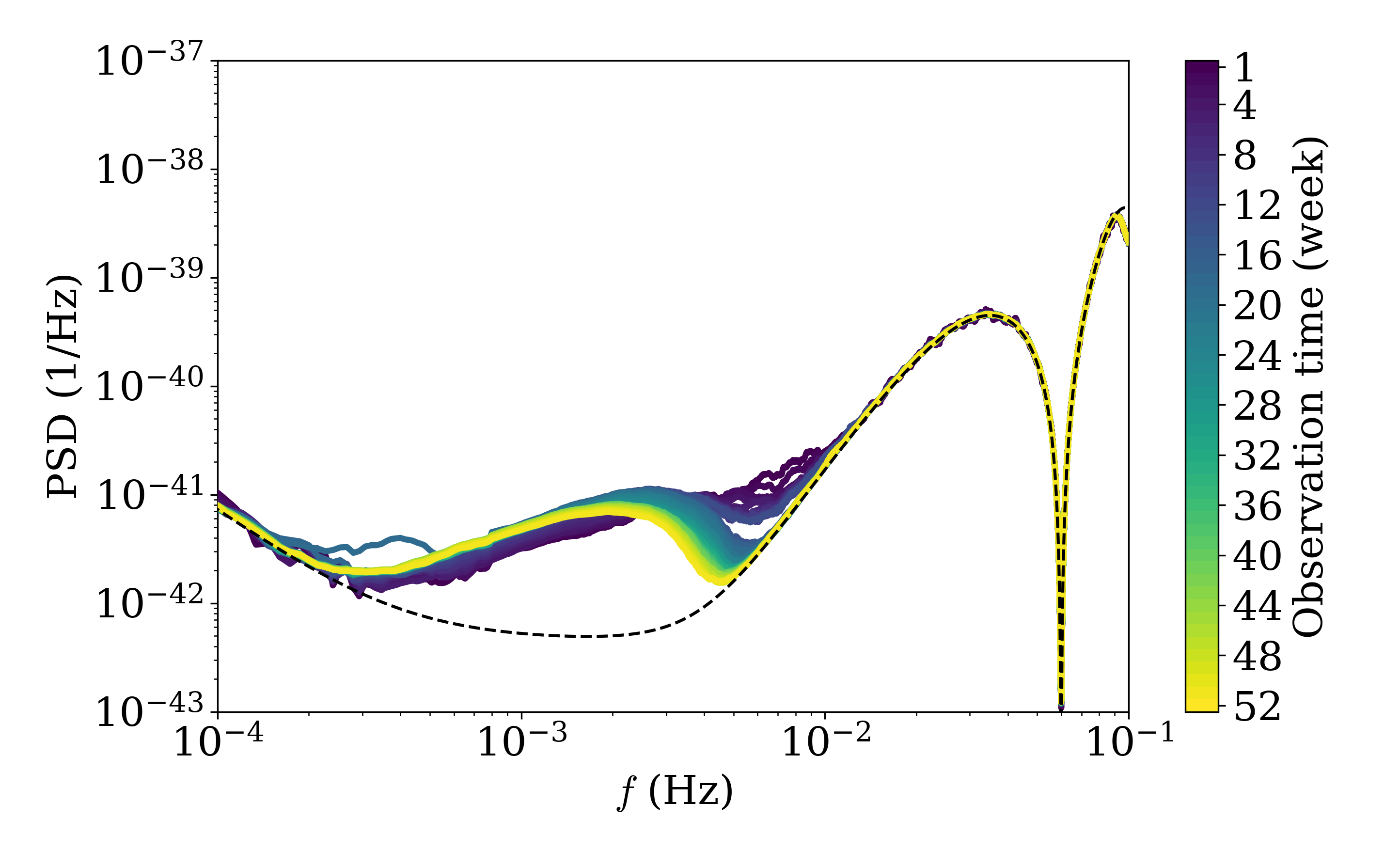}
\caption{Weekly noise estimate of the residual of the analysis of the LDC2a data set. The outlier between 0.3 and 0.4 mHz is due to a MBHB merger in week 20 whose inspiral phase is already observable in week 19 (see text). The noise estimate of each week can be accessed at \cite{strub_noise_visualizations}.}
\label{fig:noise_weekly}
\end{figure}

In Fig. \ref{fig:noise_weekly} we show the noise estimate obtained as described in Sec. \ref{sec:noise} for the weekly analysis of the LDC2a data set. The GBs and MBHBs found in the previous week are subtracted from the data. The noise estimate then consists of an estimate for each channel $n = \{ S_{A, residual}, S_{E, residual}, S_{T, residual} \}$. For $T_\text{obs} < \SI{5}{weeks} $ the noise curve is approximately linear for $f \in [ \SI{0.3}{mHz}, \SI{10}{mHz} ]$ where the detectable GBs are expected. Since the GBs are quai-monochromatic their signal is constantly recorded by LISA. On one hand, the noise estimate for $f \in [ \SI{0.3}{mHz}, \SI{2.5}{mHz} ]$ becomes louder in the first $\SI{\sim 20}{weeks}$ due to non-stationarity of the galactic confusion foreground. On the other hand, in the frequency range $f \in [ \SI{2.5}{mHz}, \SI{10}{mHz} ]$, there are fewer overlapping signals. As a result, the GB extraction algorithm can progressively analyze and subtract the GBs with greater accuracy as the observation time $T_\text{obs}$ increases.

It is also interesting to mention the increased noise at $f \in [ \SI{0.2}{mHz}, \SI{0.5}{mHz} ]$ only for $T_\text{obs} = \SI{19}{week}$. This increase is due to the MBHB merging at $T_\text{obs} \approx \SI{19.06}{week}$ which is not extracted yet but the inspiral phase is already well detectable.

\subsection{Computational Cost}
\begin{figure}[!htbp]
\includegraphics[width=0.5\textwidth]{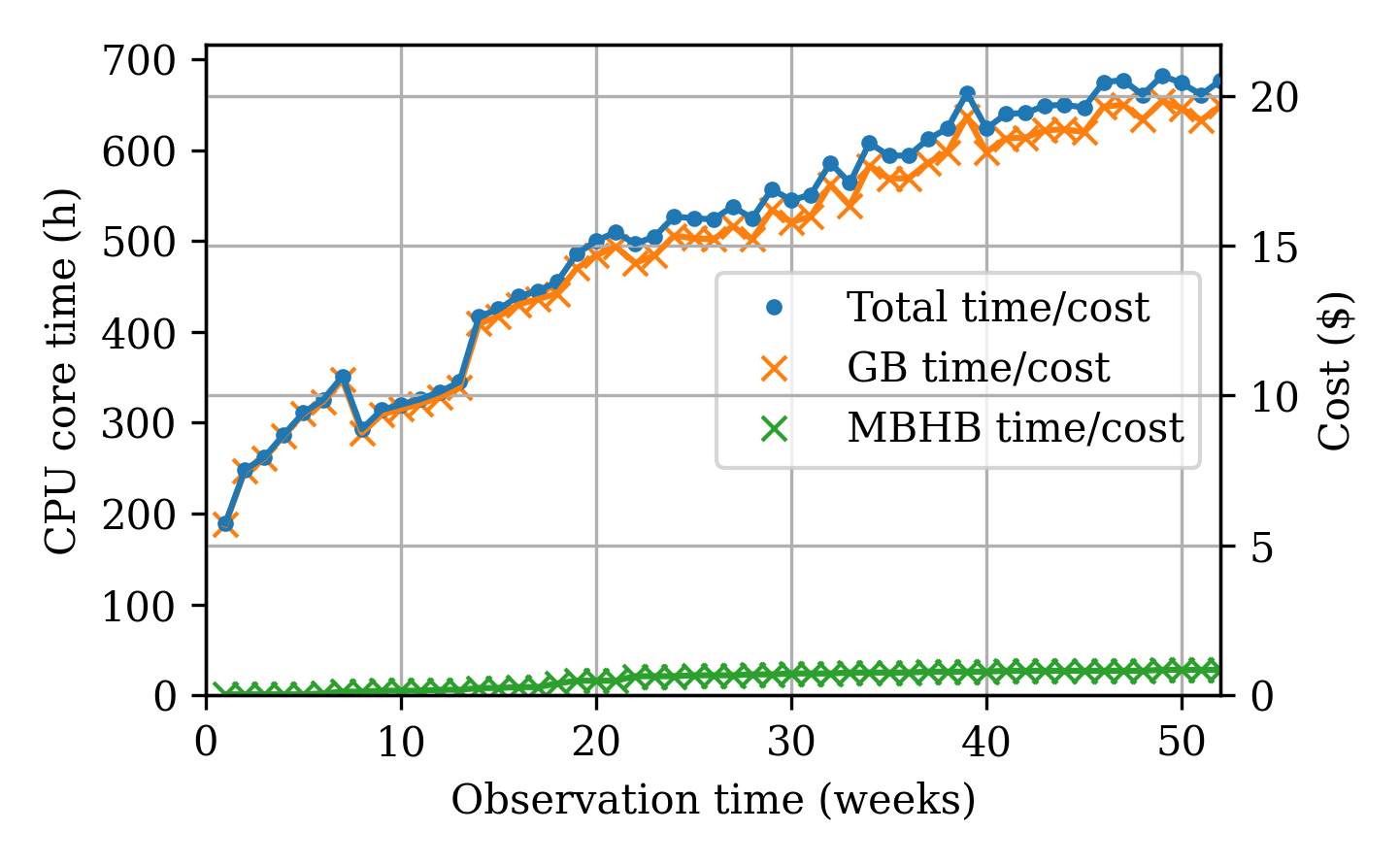}
\caption{Computational CPU core time and the correlated cost in USD if run on commercial hardware \cite{Google}.}
\label{fig:weekly_results_time} 
\end{figure}

In Fig. \ref{fig:weekly_results_time} we show the individual time and costs for each analysis as well as the cumulative sum. As shown, our pipeline can keep the computational cost at a neglectable level, which allows us to perform a weekly analysis with simulated LISA data without spending much resources.

The jump from 7 to 8 weeks is due to a change in the search repetition for each GB signal from 3 to 2, which lowers the computational cost. The jump from week 13 to week 14 is due to the change in signal generation of FastGB. At $T_\text{obs} < \SI{3}{\text{months}}$ the implemented determination of the number of frequency bins of a signal fails. Therefore we set the number of frequency bins to 128 for all signals with $T_\text{obs} < \SI{3}{\text{months}}$, which appears to be more cost-effective.

The CPU core times shown are for cost visualization only. The computation is highly parallelizable since all frequency segments can be divided into two groups of non-neighboring segments, each segment within a group can be analyzed in parallel. This brings the time for the analysis down to minutes if $\approx 3000$ CPU cores are available. For the interested reader the parallelization is discussed in more detail in \cite{strub2023accelerating}.

It is very unlikely that a new signal with $f > \SI{30}{mHz}$ becomes detectable if the data has already been analyzed for $T_\text{obs} > \SI{10}{weeks}$. Thus, to further reduce the costs one could limit the search by an upper frequency bound other than the maximum determined by $f_\text{Nyquist}$. Furthermore, the number of searches for each GB can also be reduced from two to only one to save computational costs.

\section{Robustness}
\label{sec:robustness}

Since the pipeline of this article has a low computational cost it is feasible to conduct some robustness tests. First, we test the pipeline for one iteration for the LDC2a data set with $T_\text{obs} = \SI{26}{weeks}$ where we repeat the analysis ten times with different seeds. In Table \ref{tab:evaluation robustness} the number of recovered and matched GBs for different seeds is presented. The number of recovered and matched GBs have a standard deviation of 29 and 25 GBs respectively. This is a rather low deviation compared to the mean of 5116 and 4342 GBs.

\begin{table}[!htbp]
\caption{Results of recovering signals from the LDC2a data set for $T_{\mathrm{obs}} = \SI{0.5}{yr}$ with different seeds. First, the MBHBs are recovered, and then the GBs. A recovered signal is matched if it is paired with an injected signal and the scaled error is $\delta < 0.3$.}
\begin{ruledtabular}
\begin{tabular}{cccc}
Seed  & Recovered & Matched & Match rate \\ \hline
1     & \num{5086} & \num{4311} & \num{85 \%} \\
2     & \num{5141} & \num{4385} & \num{85 \%} \\
3     & \num{5156} & \num{4361} & \num{85 \%} \\
4     & \num{5126} & \num{4359} & \num{85 \%} \\
5     & \num{5120} & \num{4327} & \num{85 \%} \\
6     & \num{5139} & \num{4340} & \num{84 \%} \\
7     & \num{5117} & \num{4364} & \num{85 \%} \\
8     & \num{5051} & \num{4299} & \num{85 \%} \\
9     & \num{5095} & \num{4327} & \num{85 \%} \\
10     & \num{5132} & \num{4351} & \num{85 \%} \\ \hline
Mean     & \num{5116.3} & \num{4342.4} & \num{85 \%} \\
Std     & \num{29} & \num{25} & \num{0.3 \%} \\
\end{tabular}
\label{tab:evaluation robustness}
\end{ruledtabular}
\end{table}

\begin{figure}[!htbp]
\includegraphics[width=0.5\textwidth]{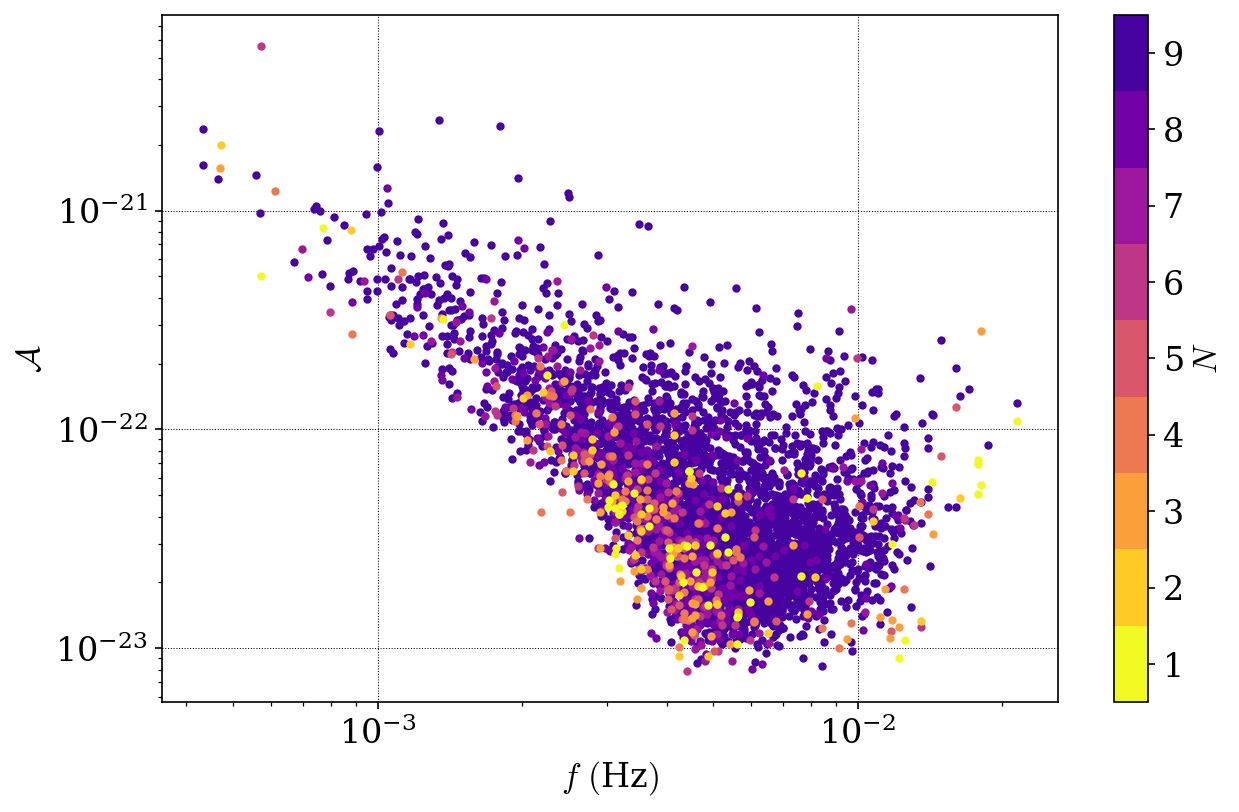}
\caption{\label{fig:robustness} Scatter plot of the recovered GBs of the analysis with seed $= 1$. The color represents the number of times $N$ this signal matches with recovered signals of the other nine analyses.}
\end{figure}

Furthermore, we matched the recovered signals of an analysis with the signals of the other nine analyses. In Table \ref{tab:evaluation seeds} we present the number of times $N$ a recovered signal is matched to the other analyses. For example, out of all the $5086$ recovered signals of the analysis with seed $= 1$ did $4184$ of them also get extracted from all nine $N = 9$ other analyses and only $N = 47$ signals were recovered from only one $N = 1$ other analysis. 
One would expect that the GBs with a higher SNR are recovered by all analyses, which is observed in Fig. \ref{fig:robustness} where most of the GBs with high amplitude are recovered by all or most of the analyses. The signal strength depends on the amplitude and the inclination angle. Therefore the signal strength does not perfectly correlate with the amplitude but the correlation is still high enough for this observation.

\begin{table*}[!htbp]
\begin{minipage}{\textwidth}
\caption{The variables of interest include the count of detectable injected GB sources, the count of recovered sources, the count of matches with injected sources, and the match rate. The match rate is determined by dividing the number of matched signals by the total number of recovered signals. The overlap is included to get an evaluation comparable to other analyses \cite{zhang2021resolving, gao2023fast, littenberg2023prototype, lackeos2023lisa, strub2023accelerating}.}
\begin{ruledtabular}
\begin{tabular*}{\textwidth}{@{\extracolsep{\fill}}cccccccc@{\extracolsep{\fill}}}
Challenge     & $T_{\mathrm{obs}}$ [yr] &  Recovered & $\delta < 0.3$ & Match rate $_{\delta < 0.3}$ & $\mathcal{O} > 0.9$ & Match rate $_{\mathcal{O} > 0.9}$\\ \hline
LDC2a     &  \num{0.5} & \num{5086} & \num{4311} & \num{85 \%} & \num{4251} & \num{84 \%}\\
LDC2a without MBHBs     &  \num{0.5} & \num{5099} & \num{4326} & \num{85 \%} & \num{4251} & \num{83 \%}\\
LDC2a     &  \num{1}  & \num{8313} & \num{7192} & \num{87 \%} & \num{7124} & \num{86 \%}\\
LDC2a without MBHBs     &  \num{1}  & \num{8323} & \num{7217} & \num{87 \%} & \num{7161} & \num{86 \%}\\
\end{tabular*}
\label{tab:sangria matches}
\end{ruledtabular}
\end{minipage}
\end{table*}

\begin{table}[!htbp]
\caption{All signal for each seed gets potentially matched with a signal from each different seed. The table shows how many signals get matched with signals from all 9 other seeds. As an example for seed = 1, $4184$ signals got matched with a signal from each other 9 catalogs of recovered signals.}
\begin{ruledtabular}
\begin{tabular}{cccccccccc}
\diagbox{Seed}{$N$} & 9  & 8 & 7 & 6  & 5 & 4 & 3 & 2 & 1 \\ \hline
1     & 4184 & 334 & 146 & 82 & 54 & 60 & 57 & 44 & 47 \\
2     & 4184 & 351 & 161 & 89 & 73 & 74 & 59 & 35 & 43 \\
3     & 4187 & 346 & 152 & 83 & 71 & 81 & 60 & 44 & 46 \\
4     & 4185 & 349 & 155 & 86 & 75 & 71 & 50 & 44 & 45 \\
5     & 4184 & 351 & 164 & 75 & 75 & 69 & 52 & 37 & 35 \\
6     & 4187 & 347 & 161 & 84 & 64 & 67 & 52 & 41 & 44 \\
7     & 4187 & 352 & 153 & 75 & 60 & 65 & 52 & 36 & 50 \\
8     & 4182 & 300 & 134 & 66 & 53 & 64 & 38 & 47 & 37 \\
9     & 4186 & 341 & 150 & 86 & 59 & 53 & 52 & 28 & 54 \\
10     & 4183 & 354 & 157 & 92 & 61 & 55 & 50 & 33 & 58 \\\hline
Mean & 4184.9 & 342.5 & 153.3 & 81.8 & 64.5 & 65.9 & 52.2 & 38.9 & 45.9 \\
Std & 1.8 & 16.1 & 8.7 & 7.7 & 8.4 & 8.5 & 6.2 & 6.0 & 7.0 \\
\end{tabular}
\label{tab:evaluation seeds}
\end{ruledtabular}
\end{table}

In Table \ref{tab:sangria matches} we present the number of recovered and matched GBs of the LDC2a data set where once the MBHBs are extracted by the pipeline and once without the MBHBs in the data. As we can see the GB extraction does not change much, which indicates a good quality MBHB extraction that is comparable with the analysis as if there were no MBHBs in the first place.

\section{Conclusion}
\label{sec:conclusion}
This article introduces a novel method for extracting MBHBs and presents a cost-effective global analysis of synthetic LISA data including MBHBs and GBs. We expand upon the GB extraction algorithm \cite{strub2023accelerating} by incorporating the MBHB extraction method, detailed in Sections \ref{sec:mbhb}, and integrating both techniques into a global fit pipeline shown in Fig. \ref{fig:flow}.

The algorithm undergoes testing on the LDC1-1 data set, featuring a solitary MBHB signal amidst instrument noise, and the LDC2a data set, comprising 15 MBHBs, 30 million GBs, and instrument noise. The MBHB algorithm effectively extracts the MBHB from the LDC1-1 data set. Furthermore, the successful extraction of self-injected signals, considering higher modes alongside instrument noise, is presented in Sec. \ref{sec:results mbhb}.

In the challenge of the LDC2a data set, the MBHB algorithm extracts MBHBs such that the GB extraction and noise estimation is without interference effectively treating obtaining a residual as if the MBHBs were initially absent from the data set. Only if a strong MBHB signal merges right at the boundary in the next week, the MBHB is visible in the noise estimate but still not interfering with the GB extraction. Future work could extend the MBHB signal extraction to include MBHBs whose mergers have not yet been measured.

Thanks to the low cost of this pipeline, where analyzing $T_\text{obs} = \SI{1}{yr}$ costs only $\SI{700}{CPU\, core\, hours}$, corresponding to $\SI{20}{USD}$, we can conduct weekly analyses as new data is incorporated. Note that 52 weekly analyses require a cumulative cost of only $\SI{800}{USD}$. Furthermore, the pipeline operates on multiple frequency segments in parallel, thereby reducing the analysis duration to mere minutes.

Starting from week 16, we observe a linear increase of 129 recovered GBs per week with a constant match ratio of $85\%$ to $88\%$. Moreover, weekly analyses reveal the correlation between $T_\text{obs}$ and the difference between recovered and injected parameters.

Moreover, the noise estimation is effectively extended to arbitrary $T_\text{obs}$. The weekly analysis unveils the evolving noise curve, showcasing enhanced GB extraction for signals with $f > \SI{2}{mHz}$ as $T_\text{obs}$ increases, aligning the noise estimate of the residual closer to the instrument noise level.

Furthermore, we leverage the low cost of the pipeline to conduct robustness tests by repeating analyses of $T_\text{obs} = \SI{0.5}{yr}$ with different seeds.

The result is a robust, computationally efficient, high-quality GB and MBHB signal extraction and noise estimation pipeline publicly available at \cite{strub_gb_parameter_estimation, strub_mbhb_parameter_estimation}.

The next steps involve incorporating other GW signal types such as extreme mass ratio inspirals, glitches, and data gaps. Additionally, the pipeline could be extended to include pre-merger MBHB parameter estimation, as the current pipeline focuses solely on post-merger MBHB signal extraction. Furthermore, the found MLEs of both MBHBs and GBs can be used as proposals of computationally expensive MCMC searches to improve the sampling efficiency. Therefore this approach could be integrated with MCMC based global fit, GB and MBHB algorithms such as \cite{littenberg2023prototype, cornish2020black, marsat2021exploring, katz2020gpu, cornish2022low, katz2022fully, PhysRevD.75.043008, crowder2007genetic, littenberg2011detection, 10.1093/biomet/82.4.711, gamerman2006markov,littenberg2020global, littenberg2023prototype, karnesis2023eryn, tong2024transdimensional} to improve the computational efficiency.

\section{Acknowledgements}
We are thankful to the LDC working group \cite{LDC} for their efforts in establishing and backing the LDC1-1 and LDC2a. Additionally, we appreciate the GPU implementation of $\texttt{FASTLISARESPONSE}$ \cite{michael_katz_2022_5867731} and the signal simulation of the MBHBs  $\texttt{BBHx}$ \cite{katz2020gpu, katz2022fully} which is available at \cite{michael_katz_2021}. We thank Franziska Riegger for her helpful feedback. We thank the referee for the thorough review which we believe improved this article. The computations to extract GBs were conducted on the CPUs of the Euler cluster at ETH Zürich, and we extend our sincere appreciation for this support. This project receives funding from the Swiss National Science Foundation (SNF 200021\textunderscore185051).

\bibliography{references}

\end{document}